# Spin Relaxation in Quasi-1D GaAs Mesowires: Control via Electric Field and Aspect Ratio

Stefania Castelletto[1], ChanJu You[2], Deborah L. Gater[3], and A. F. Isakovic[2,*]

[1]RMIT, Melbourne, Australia

[2]Colgate University, Hamilton, NY, USA

[3]University College London, London, UK

**Abstract:** We report on the measurements of spin relaxation in GaAs quasi-one-dimensional mesowires, relying on spin noise spectroscopy, thus adding to the existing body of spin relaxation studies in bulk, two-dimensional and zero dimensional systems. In addition to temperature and magnetic field dependence, we modify the spin relaxation time via applied electric field and aspect ratio of the mesowires, suggesting that scalable spintronics devices with controllable spin relaxation are achievable. Overall, we observed higher spin-relaxation time in mesowires compared to bulk with a spin noise exhibiting D'yakonov-Perel scattering and other scattering behavior. Spectral spin noise data are interpreted in part via Glazov-Sherman model [1], where both, diffusive and ballistic spin relaxation are accounted for.

(*) corresponding author: aisakovic@colgate.edu; iregx137@gmail.com

**Introduction and Motivation**

The use of electron spins as qubits for quantum computing among other applications remains an area of active research, particularly in the field of spintronics – a field of physics that studies spin dynamics and spin-dependent phenomena via spin noise spectroscopy in information processing [2].

Spintronic devices exploit electronic currents that are spin polarized, having an excess of one spin states over the other. Approximately two decades ago, the field of semiconductor spintronics started being of major interest due to the advantages over charge only based devices such as the potential to surpass the limits of speed-up and power consumption [3]. Spin- and light-polarization effects in bulk and nanostructured semiconductors, where photon and charge confinement occur, are therefore at the core of the implementation of innovative optoelectronic devices for spintronics.

Relatively early on, the community has realized two scale-down obstacles: (a) the spin phenomena we study are not always by default quantum spin phenomena, and (b) spin noise is ubiquitous. With a few exceptions, the work on spin noise to date has focused on studying spin noise in materials, such as bulk with various doping conditions (3D) [4, 5], quantum wells (2D) [6-13], or quantum dots (quasi-0D) [14-16]. This is understandable, given that these materials are more accessible than (nano)devices, so this work reports on simple (2-contacts) device configuration, and on quasi-1D (mesowire) geometry, which has also rarely been studied compared to other dimensionalities.

One of the main aims of spintronics is the optimization of performance through maximizing the spin-decoherence or spin relaxation time [17]. Spin-dephasing related to spin-orbit interaction is however the main mechanism limiting the electron spin lifetime in III-V semiconductors and the spin-transport in spintronics devices based on these semiconductors [18]. Spin-dephasing mechanisms in bulk semiconductors and their nanostructures have been studied for many years with various methods [19, 20], however there is not yet a definitive description of spin-dephasing mechanisms in spintronics devices. There is a particular need to understand the effect of sample geometry on spin noise and thus spin lifetime (i.e. bulk 3D vs quantum well 2D, vs meso or nanowire (quasi-)1D vs quantum dot 0D), as different geometries are relevant for different device configurations and therefore for different applications [19, 20].

Spin relaxation is closely related to spin noise – spin fluctuations – as noise gives information on spin dynamics under external perturbations [20], and thus many studies on spin noise in various systems have been done under varying conditions such as temperature, electric bias, magnetic field, sample geometry, and doping [4, 5, 21, 22]. Spin noise measurements provide information about spin relaxation time (spin dephasing time), $\tau_s$, its power spectrum and the electrons' or holes' Landé g factor.

Spin noise measurements in semiconductors can be performed by several methods ranging from electron paramagnetic resonance (EPR) [23] and nuclear magnetic resonance (NMR) [24], magneto-optical measurements (time-resolved photoluminescence, Hanle effect, Kerr rotation microscopy, Faraday's effect [22, 25] and magnetic force microscopy [26]. A particularly useful technique is spin noise spectroscopy (SNS), which consists of optical measurements of spin noise via Faraday rotation of a probe light, induced by the stochastic spin polarization of an electron/hole ensemble. One advantage of using SNS is that it dissipates (almost) no energy in the sample, thus leaving the spin dynamics undisturbed when taking spin measurements [27] (For example, a polarized laser probe with energy slightly off resonance relative to the energy bandgap of the material introduces spin–orbit coupling between carriers and the laser photons helicity via dipole selection rules. The carriers' ensemble spins are mapped onto the transmitted light polarization via the Faraday's effect. The method is known in atom optics and has been adopted to study spin noise for spintronics material and devices [14, 28].

Several factors contribute to the large variability of spin noise measurements even in the same material, including: doping concentration, magnetic field, temperature, probe laser wavelength and intensity, and sample volume, and material dimensionality. A key challenge to determine the stochastic electron spin fluctuations at thermal equilibrium is to maintain nominally "un-perturbative" measurement conditions, consisting in avoiding optical absorption of the laser probe, maintaining low intensities and large spot sizes.

Typically, SNS methods in doping conditions below the metal to insulator transition, provide a lower bound in spin dephasing time values compared to other methods, where spin dephasing dependence on laser intensity and magnetic field are very relevant. Many of the previous studies of spin noise have been performed in III-V GaAs based semiconductors systems due to its controllable purity and easy availability [20, 29, 30], although other materials have also been investigated, including ZnO [31], Mn ions diluted in CdTe [32] and electron-doped monolayers of $MoS_2$ [33], to name but few.

One aim of spin noise measurements is to elucidate the dominant dephasing mechanisms in different conditions (temperature, material, geometry, etc.). The major identified spin dephasing mechanisms in semiconductors are summarized in Table 1. These mechanisms are present with different contributions across different dimensionality as well.

*Table I A brief review of main spin relaxation processes in semiconductors*

| Spin Relaxation Type | Physical Situation | Relaxation Characteristics |
|---|---|---|
| Elliott-Yafet (EY) | • Momentum scattering<br>• Phonons, impurities<br>• Small band gap | $\tau_s \propto T^{-1/2}$ |
| D'yakonov-Perel (DP) | • Spin-orbit splitting of conduction band w/ lack of inversion symmetry<br>• n-doping | $\tau_s \propto T^{-\frac{3}{2}}$<br>$\tau_s \propto T^{-3}$ |
| Bir-Aronov-Pikus (BAP) | • Spin flipping<br>• Overlap e⁻ and h⁺ wavefunctions<br>p-doping needed | $\tau_s \propto T^{-1}$ |
| Hyperfine relaxation | • Due to hyperfine interaction<br>• Overlap e⁻ and (nucleus) N⁺ wavefunctions<br>• Ubiquitous at low temperatures (70K max) | $\tau_s \propto T^{-n}$<br>$n$ – experiment dependent |

We will draw conclusions on spin noise data in 1D-like fabricated GaAs nanostructures by reviewing the current understanding of spin noise from bulk 3D to 2D and 0D dimensionality, in regard to the interpretation of the noise mechanism.

Spin relaxation time of bulk GaAs (n-doped or undoped) has been intensively studied by various methods and ultimately by SNS to determine its temperature, doping and magnetic field dependence, resulting now in a relatively well understood picture of the main dephasing mechanisms, that occur depending on doping and temperature up to room temperature. The dephasing mechanisms observed in bulk materials are associated with observed spin relaxation time dependence on doping and temperature.

In Fig. 1(a) we summarize several published measurements of the spin dephasing time, $\tau_s$, in bulk GaAs for different doping concentrations and temperatures. The spin noise measurements were done using different methods as described in the references, with doping variations studied over 3 orders of magnitude. Based on doping dependence, three regimes could be identified insulator-like, metal-like and the metal to insulator region (intermediate). In the metal-like phase, electrons are delocalized so the D'yakonov-Perel (DP) [34] mechanisms are dominating, providing spin dephasing time below the ns, while in the insulator phase, spin dephasing is due hyperfine interaction, permitting to achieve above 100 ns dephasing time.

For bulk GaAs, temperature effects on the spin relaxation time for all doping conditions are weak up to 10K, where the electrons ensemble can be considered localized. For temperature above 10K up to 80K, for low doping and mixed phase doping below the metal-insulator phase, $\tau_s \propto T^{-1.48}$ (i.e. DP mechanism). In the metallic phase more pronounced temperature dependence was observed with $\tau_s \propto T^{-1.96}$, mostly due to scattering at the ionized impurities (EY mechanism). Closer to room temperature for dopant concentration from 2-3x10$^{16}$ cm$^{-3}$, $\tau_s \propto T^{-3}$ [5, 35], following pure DP dephasing mechanism [4, 36].

The DP mechanism can be suppressed by dimensionality reduction form 3D→2D. In 2D structures, DP dephasing arises from two main contributions: intrinsic bulk inversion asymmetry (BIA) and structure induced asymmetry (SIA). The second can be controlled by the application of an electric field in specific growth direction, thus permitting at a critical field to compensate for the BIA dephasing term and thus suppress the DP mechanism [6].

For example, QWs (quantum wells) in GaAs attracted attention for spintronics due to their potential engineering of their growth direction along unconventional crystallographic axes to suppress DP effects due to the special symmetries of the SOI in these structures. Recent SNS measurements on one hand confirmed the suppression of DP-related spin dephasing mechanism. However other dephasing mechanisms have been observed related to intrinsic spin noise fluctuations due to the larger extent of localization of carriers and optical creation of holes, which lead to the decay of electron spin via the Bir-Aronov-Pikus (BAP) [37] mechanism and recombination with spin-polarized electrons. Studies at room and low temperature and in the presence of B-field have also been performed providing methods to achieve above 100ns (up to 300ns) spin dephasing time by controlling spin orbit symmetry effects in (110) and (111) grown QWs. The QW spin dephasing time at low temperature can be controlled by either reducing optical excitation or by applying electric field and surface acoustic waves [10].

The relatively recent model of spin dephasing in QWs provides a spin decay rate given by [38, 39]:

$$\frac{1}{\tau_s} = \frac{1}{\tau_s^{lim}} + \gamma_s^{BAP} N_h + \gamma^r N_h \quad \text{Eq. 1}$$

where $\frac{1}{\tau_s^{lim}}$ is the spin dephasing in the limit of zero excitation. In this regime spin dephasing is due to either EY or to DP mechanism. The terms $\gamma_s^{BAP} N_h$ and $\gamma^r N_h$ describe spin dephasing due to BAP mechanism, where $N_h$ is the steady state holes density and $\gamma^r$ is the recombination rate of polarized electrons with holes.

In Fig. 1(b) we summarize measured dephasing time InGaAs QWs grown on (110) and (111) GaAs, versus doping concentration and temperature. We note more than 2 orders of magnitude in doping variation, and a wide temperature range, including above 300 K. A less dramatic drop in $\tau_s$ as T → 300 K is observed, with relaxation time from 100 ps to 300 ns, which provides technologically relevant relaxation time, while moderate dependence on magnetic field. Dephasing via the SOI can be further suppressed in GaAs (111) QWs by applying an electric field [6, 7, 10], obtaining tunable spin lifetime [7].

Due to the high localization and dilute ensemble of spin in quantum dots (QDs), the DP mechanism is unlikely to be a contributor to dephasing as it is in 2D systems. Therefore, QDs appear very attractive for quantum spintronics. Many spin noise measurements have been performed in QDs from ensemble to single spin at low temperature, recording very long relaxation time for holes spin, as QDs are unintentionally p-doped.

Notably, SNS was performed in ensemble of $10^{10}$ cm$^{-2}$ self-assembled In(Ga)As/GaAs QDs per layer, grown on (100) GaAs, nominally undoped ($10^9$ holes cm$^{-2}$) [14], indicating the practicability of the method to study intrinsic spin dynamics using non-demolition methods. It was found that electron and hole spin fluctuations generate distinct noise peaks; the dephasing rate, increases linearly with B, yielding an inhomogeneous relaxation time for hole of 1ns at B = 0.1 T at 6 K due to an inhomogeneous distribution of holes *g* factor.

In Fig. 1(c) we summarize of measurement of spin dephasing time using various methods in QDs, providing a relaxation time spans of 5 orders of magnitude (up to 100 μs), spin noise controlled by magnetic field, operation only at 4K. Recent studies on spin noise and spin relaxation time in semiconductors have provided valuable insight into their potential for exploring various experimental conditions such as temperature [40], applied local magnetic field [41], pump power density [42], wire diameter [43], and the piezoelectric field [44].

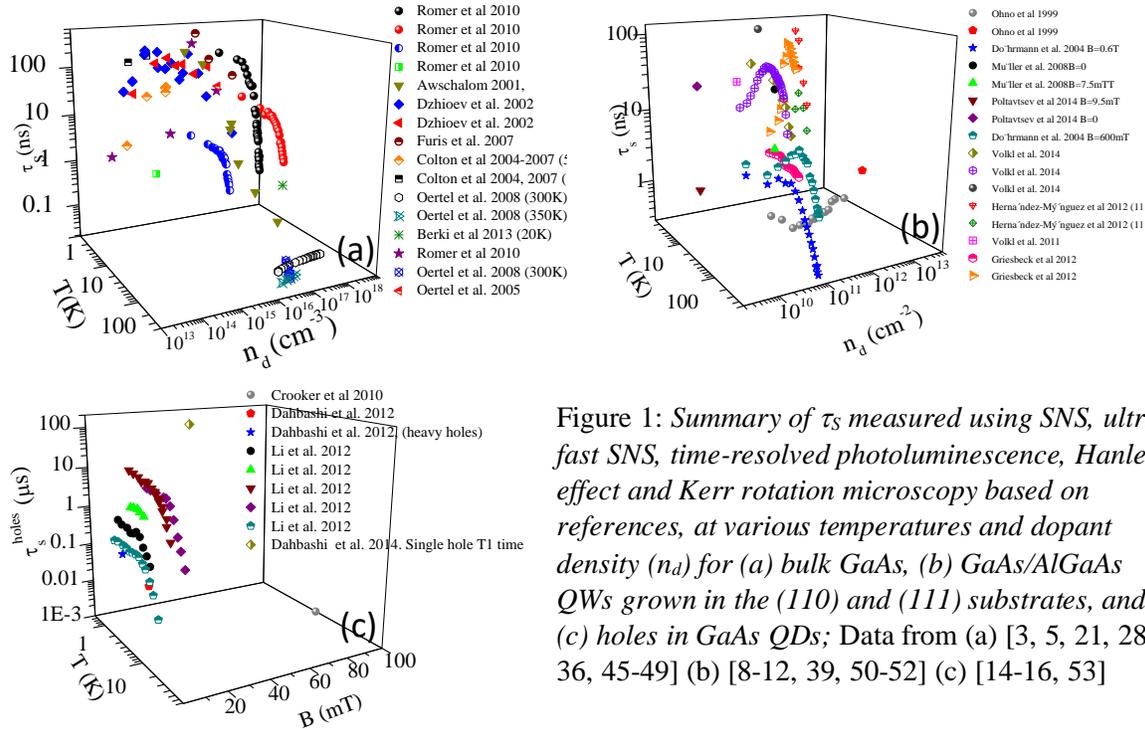

Figure 1: *Summary of $\tau_S$ measured using SNS, ultra-fast SNS, time-resolved photoluminescence, Hanle effect and Kerr rotation microscopy based on references, at various temperatures and dopant density ($n_d$) for (a) bulk GaAs, (b) GaAs/AlGaAs QWs grown in the (110) and (111) substrates, and (c) holes in GaAs QDs;* Data from (a) [3, 5, 21, 28, 36, 45-49] (b) [8-12, 39, 50-52] (c) [14-16, 53]

**Conclusions Drawn from Review**

A number of clear conclusions and reasonable inferences might be drawn from the admittedly brief review we offer in the previous section. Among clear summary points are:

*Conclusions and Observations*

a) Spin relaxation via spin noise has been explored in most of the technologically relevant doping ranges in both 3D and 2D.
b) Very broadly, one can identity spin relaxation times across the range of $\tau_S$ [100 ps, 300 ns] and over 4 orders of magnitude in doping $n_d$ [$10^{14}$, $10^{18}$cm$^{-3}$], and in temperature range between 1.5 K and 350 K, for 3D.
c) Spin relaxation has been studied over the span of three orders of magnitude in 2D, including technologically relevant near room temperature conditions.

d) In both, 3D and 2D, magnetic fields used in some of the experimental setups are close to and in the technologically relevant range, in that it is easier to implement $10^{-3}$ T range and below technologically, on future chips that would rely on semiconductor spin signal processing and/or storage.
e) The range of spin relaxation time values in all three dimensionalities is at least 4, often 5 orders of magnitude.

*Inferences and Conjectures*

a) The change of the spin relaxation time with temperature appears to be a different analytical expression for different dimensionalities.
b) If it is possible to obtain higher spin relaxation time in lower dimensionalities, as the 0D summary would indicate, it is reasonable to examine what one can obtain for the spin relaxation time in 1D.

**Samples and Experimental Setup**

This work is motivated by the desire to enable, examine, and understand spin noise phenomena in samples of (quasi-)1D dimensionality. There are essentially two ways one can proceed – either by growing semiconductor nanowires, as some recently shown in GaN nanowires [54], GaAs/AlGaAs core/shell [55] or by nanofabricating the wires out of the substrate. We have opted for the latter approach, as it allows us to study spin noise with a focus on meso- and nanodevices applied electric field. The nanodevice focus is necessary for several reasons, a chief among them being the need for realistic estimates of the applicability of spintronics devices in reduced dimensionalities, with an eye towards spin noise control. Our goal was to study quasi-1D mesowires in an as isolated state as possible, and we therefore aimed to make standalone wires with some minimal support. The most stable configuration we fabricated is shown in Fig. 2, where we show (not to scale) an array of wires over significantly thinned down undoped bulk semiconductor substrate. GaAs [100] substrate is nominally undoped (very lightly n-doped) whereas the thin film of grown GaAs was doped to $7 \times 10^{16}$ cm$^{-3}$. The thinning down of the substrate supporting side is accomplished with a combination of deep reactive ion etching (DRIE) and spray etching, after the supporting structure is protected by the photoresist. The wire array and its accompanying contacts are fabricated on top and protected before performing the DRIE etching. The contacts, made by depositing Ti-Au bilayer structure, allow for the application of external electric field, so that spin noise in these wires could be studied as a function of applied bias.

Wires of various aspect ratios were fabricated, and two examples are shown in Fig. 3a and 3b. We looked at several lateral aspect ratios $a \times b$ ranging from 80 nm × 30 nm to 25 nm × 125 nm, with the length in the order of 5-10 μm. The back-etched samples allowed back access with infra-red light after arrays of wires were nanofabricated on the front with electron beam lithography. One face of the rectangular mesowire is still in contact with the substrate, which is a limitation.

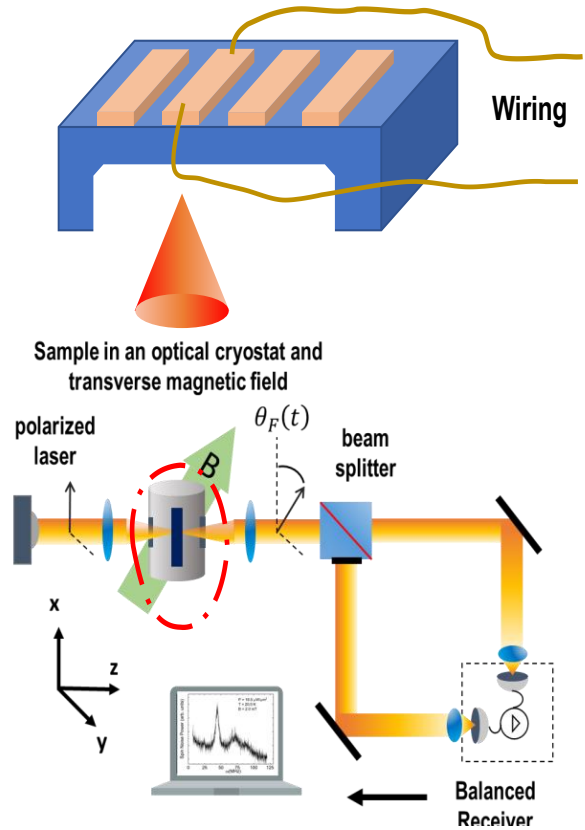

*Figure 2: (top) A schematic of the nanofabricated wires on a supporting semiconductor substrate, where much of the substrate is removed to allow for the easier light access. (bottom) a sketch of the standard experimental setup for Spin Noise Spectroscopy, with device from (top) being in cryostat.*

Laser beam with wavelength of 820 nm and higher is microfocused onto a sample with a high resolution thanks to an oil coated microlens placed in front of the cryostat window. A linear polarization ratio is kept at $10^{-5}$ by a combination of a linear polarizer and a spatial filter. The magnetic field is applied perpendicular to the laser beam, and its strength is measured by a Hall probe placed inside the cryostat close to the sample mount. Data are collected by parallel scanning of the spectrum and the real FFT processing.

The beam location over mesowire is established by calibrating nanopositioning system to spatially dependent peaks in photovoltage, the latter being recorded via contacts to the mesowires. For calibration purposes, we plot photovoltage as a function of lateral shift of nanopositioning system, where peaks in photovoltage spatially match distances between nanofabricated mesowires.

To study spin relaxation processes under the influence of applied electric field, contact pads (Fig. 3c) were fabricated typically in side-geometry on 10 μm scale, to allow for a convenient wire-bonding and wiring connection from inside the cryostat to outside transport electronics.

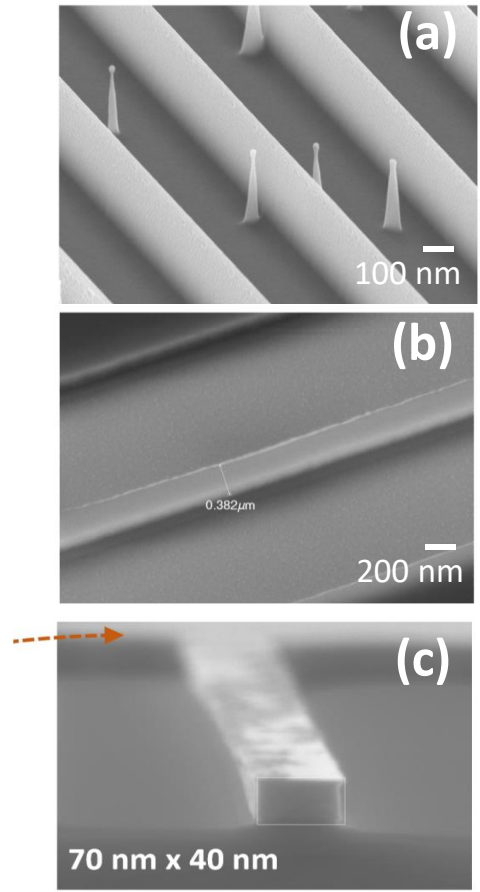

*Figure 3: Panels (a) and (b) show mesowires with varied aspect ratios. (c) a side view of the contact pad (back), indicated by the arrow, connected to the nanowires (front).*

**Experimental Results and Discussion**

A typical spectrum obtained in our setup for the nanowires is shown in Fig. 4a. However, we note a couple points: first, it seems that the raw spectrum "sits" on top of a $\frac{1}{f}$ - like spectrum, which is likely the consequence of the quasi-1D sample geometry, particularly as similar features are not observed in spin noise spectra for 2D or 3D samples, to the best of our knowledge. The second unusual feature, which distinguishes spin noise spectra of mesowires from those of 2D and 3D samples, is the appearance of a secondary, broader peak at a slightly higher frequency (~70 MHz). We find an approximately linear trend of the main spin noise spectra peak with respect to the B-field (Fig. 4b). The uncertainties in the peak position are determined from Lorentzian fits, and it is understandable that the uncertainties are larger for higher temperatures. Relatively small uncertainties in the values of the magnetic field originate from the measurement of the magnetic field close to the sample by a calibrated Hall-bar.

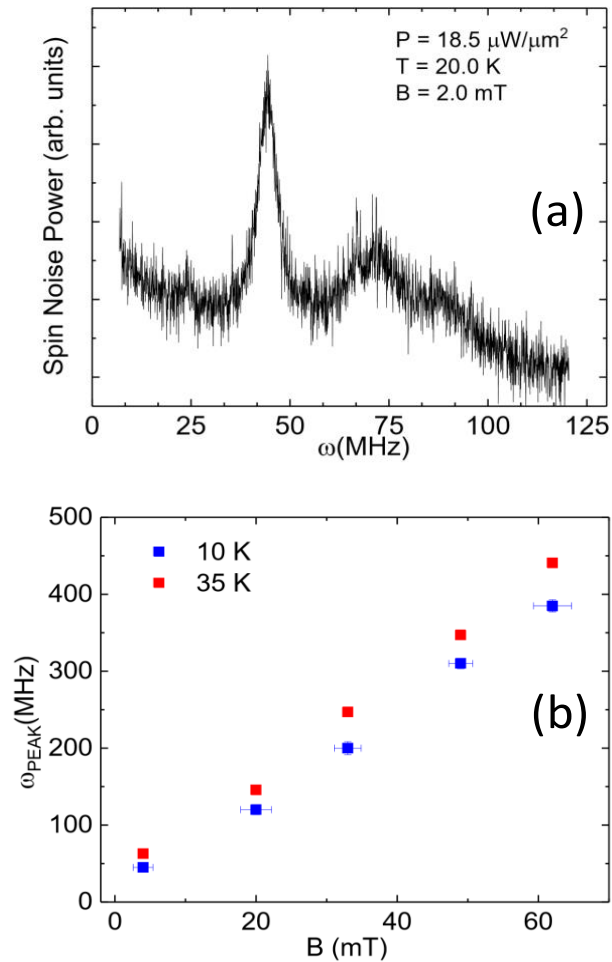

*Figure 4: (a) Example of spin noise spectra under the experimental conditions of this experiment. (b) Dependence of the peak spin noise frequency on the external magnetic field.*

The main focus of the study is to examine spin relaxation time as a function of temperature, applied voltage, and the wire aspect ratio. In what follows, we rely on the analysis of the main dominant peak from Fig. 4a, whereas the possible origin and discussion of the secondary peak (~70 MHz) will be subject discussed with Fig. 6.

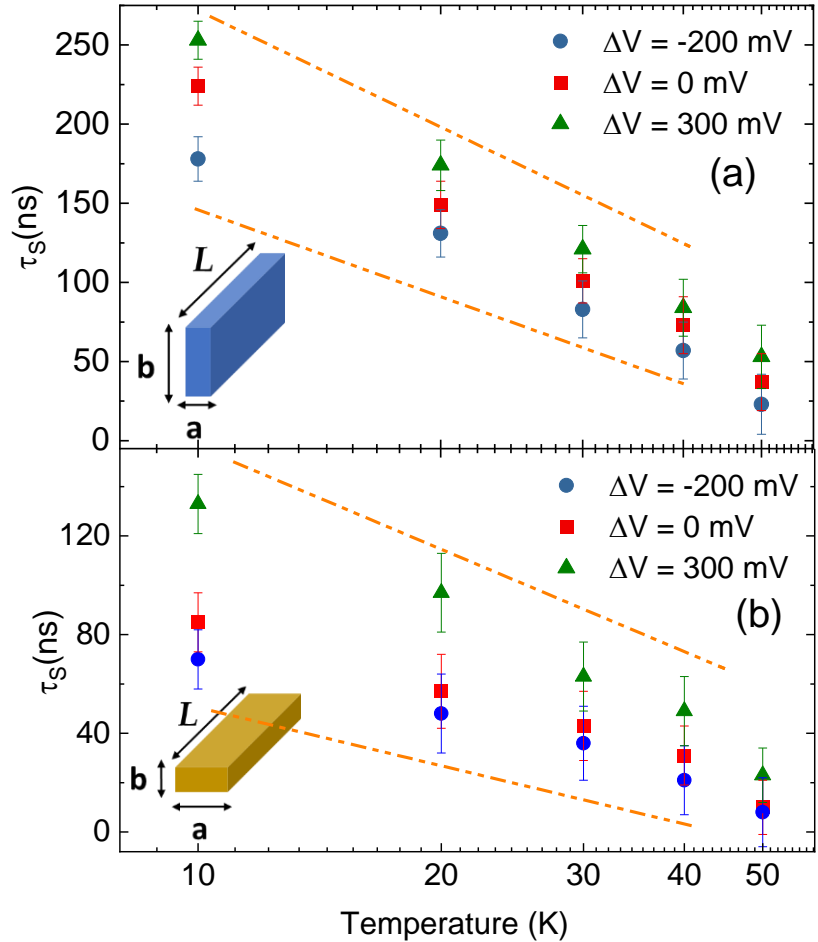

*Figure 5: Dependence of relaxation time on temperature for different electric bias values. Panels (a) and (b) represent two different aspect ratios of the mesowires, where the rectangular shapes are indicative of the aspect ratio of the mesowires. (a) a:b = 30 nm:100 nm; (b) a:b = 100 nm:30 nm. The lines are only a guide to the eye. The quasi-1D mesowires are highly anisotropic as L ≫ a, b.* b ∥ [100] GaAs

In comparison of Fig. 5a and 5b, we see some apparent similarities such as:

1) There is an inverse proportionality of spin relaxation and temperature for different aspect ratios and for a variety of biasing conditions.

2) There is a noticeable change of spin relaxation time of approximately a factor of 2 between two main aspect ratios we report on. Namely, the spin relaxation time increases for the wires where the height to width aspect ratio is greater than 1, possibly due to a reduced contribution to spin noise of the interface with the substrate. As a reminder, "height" is measured along the [100] direction.

    3) We also note that spin relaxation time could be controlled by changing voltage polarity and by voltage amplitude by a factor of approximately 15-30%, depending on temperature. The observed temperature dependence of the spin relaxation time could be partly attributed to DP mechanism due to the conduction-band electron spins (not localized electrons) in n-GaAs, but we note that analysis based on the model $\tau_s = \beta\, T^{-k}$ [56], as prescribed in Table 1, has validity in limited range of temperatures in our data. Furthermore, we see that the slope of the linearized version of the preceding formula, $\log \tau_s = \log \beta - kT$ varies with amplitude and polarity of the bias. This finding points towards quantifiable change of spin relaxation time with variations of applied voltage and aspect ratios of the quasi-1D wires. Because we focused on the aspect ratio and bias influence, this report shows a relative scarcity of temperature dependent spin relaxation data and therefore we are reluctant to propose a model at this stage.

The effect of the electric field (applied voltage) on the spin relaxation time is likely mainly influenced by the changes in the kinetic energy of spin polarized electrons. The changes due to aspect ratio are likely in part due to modifications of the lateral confinement of carriers.

Glazov and Sherman [1] propose a model of spin relaxation in nanowires which proceeds in two regimes, diffusive and ballistic. Diffusive is characterized in a manner similar to spin relaxation in other dimensionalities, whereas ballistic appears under some conditions in nanowire geometry. We hypothesize that the overall spin relaxation process we observe here might be a linear combination of both contributions. To this end, in addition to a previously known line shape,

$$\left\langle S_z^2 \right\rangle_\omega = \frac{2\tau_{S,DIFF}}{1+\omega^2 \tau_{S,DIFF}^2} \qquad \text{Eq. 2}$$

we used the following expression, derived in Glazov-Sherman paper [1].

$$\left\langle S_z^2 \right\rangle_{\omega,BAL} = 2\tau_{S,R}\, \text{Re}\left( \frac{1-i\omega\tau_{S,R}}{\omega_0^2 \tau_{S,R}^2 + \left(1-i\omega\tau_{S,R}\right)^2} \right) \qquad \text{Eq.3}$$

to model the behavior of the spin noise power spectrum with weighted, linear combination of two processes. We propose that the two peaks will be a linear combination of two spin noise power functions (Eq. 2, 3), each with different ω₀ values and coefficients $\alpha$ and $\beta$. The output is shown in Fig.6. Similarity to data presented in Fig. 4a is clear in the presence and variability of the secondary peak.

A limitation of our approach is that the mesowires we nanofabricated may not satisfy all of the "ballistic spin" conditions assumed in derivation by Glazov and Sherman.

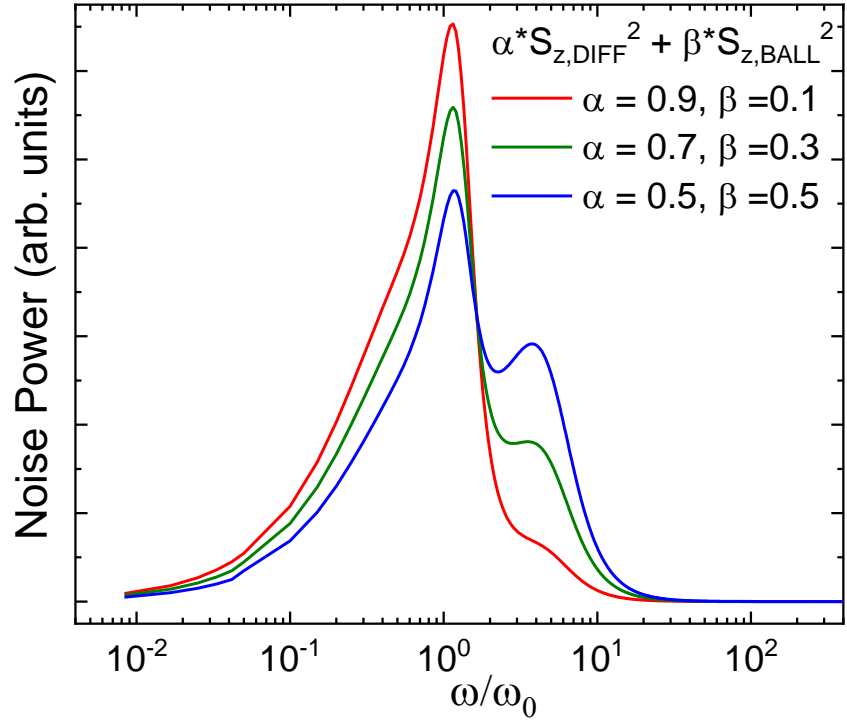

*Figure 6: A weighted linear combination of the diffusive and ballistic regimes from the Glazov-Sherman model predicts a two-peak SNS spectrum. $\omega_0 = 45.2$ MHz*

**Conclusions and Open Questions**

Motivated by the relative scarcity of spin noise spectroscopy studies in quasi-1D systems, we fabricated GaAs mesoscopic wires, and contacts for electric bias, and prepared samples where the quasi-1D nature is emphasized through careful nanofabrication processes. Depending on the aspect ratio of the cross section of the mesowires, we see spin relaxation time in the range of 25 - 300ns, and we see that application of voltage in the ± 300 mV can change relaxation time by ± 25-40% at any fixed temperature below 50K. Analysis of temperature dependence of the spin relaxation time points towards the existence of two relaxation regimes, one of which is D'yakonov-Perel' – like, while the other regime(s) may require additional studies. Additionally, the SNS spectra for quasi-1D GaAs mesowires show secondary SNS peak. This dual peak structure could be explained by a weighted linear combination model based on the Glazov-Sherman theory of spin relaxation in nanowires.

## Acknowledgments

This work has been supported in part by Mubadala-SRC grant 2013-VJ-2335, and in part by 2016 ADEC-A2RE grant. A part of the work was done at Cornell University Center for Nanofabrication, supported by the National Science Foundation under Grant No. NNCI-2025233. A part of this project used resources of the Center for Functional Nanomaterials, which is a U.S. DOE Office of Science Facility, at Brookhaven National Laboratory under Contract No. DE-SC0012704. Initial work on the project that led to this manuscript started when DLG and AFI were at Khalifa University of Science and Technology (KUST), Abu Dhabi, UAE. SC acknowledges hospitality of KUST. CY acknowledges the support of Colgate Undergraduate Research Fund. AFI acknowledges technical assistance from Dr. C. Alpha (Cornell University, CNF) and Dr. A. Stein (Brookhaven National Laboratory, CFN), and hospitality of Cornell University during parts of this project.

## Data availability

Data used to prepare this manuscript are available upon reasonable request.